\UseRawInputEncoding

\documentclass[aps,prl,twocolumn]{revtex4-2}

\usepackage[T1]{fontenc}
\usepackage{amsmath}
\usepackage{amsfonts}
\usepackage{amssymb}
\usepackage[english]{babel}
\usepackage{hyperref}
\usepackage{graphicx}
\usepackage{gensymb}
\usepackage[version=4]{mhchem}
\usepackage[export]{adjustbox}
\usepackage{array} % for m{} column specifier with vertical centering
\usepackage{float}
\usepackage{bm}

\usepackage{xcolor}

\graphicspath{{./}{Figures/}}

\begin{document}
\title{{Transfer Learning for Analysis of Collective and Non-Collective\\Thomson Scattering Spectra}}

\author{T. Van Hoomissen$^{1}$}
\author{J. Alhuthali$^{1}$}
\author{A.M. Ortiz$^{1}$}
\author{D.A. Mariscal$^{2}$}
\author{R.S. Dorst$^{1}$}
\author{S. Eisenbach$^{1}$}
\author{H. Zhang$^{1}$}
\author{J.J. Pilgram$^{1}$}
\author{C.G. Constantin$^{1}$}
\author{L. Rovige$^{1}$}
\author{C. Niemann$^{1}$}
\author{D.B. Schaeffer$^{1,}$}%
 \email{dschaeffer@physics.ucla.edu}

\affiliation{%
\vspace{0.22cm}
$^{1}$Department of Physics and Astronomy, University of California - Los Angeles, Los Angeles, CA 90095, USA \\
$^{2}$Lawrence Livermore National Laboratory, Livermore, CA 94550, USA}

%\date{\today}

\begin{abstract}
Thomson scattering (TS) diagnostics provide reliable, minimally perturbative measurements of fundamental plasma parameters, such as electron density ($n_e$) and electron temperature ($T_e$).  Deep neural networks can provide accurate estimates of $n_e$ and $T_e$ when conventional fitting algorithms may fail, such as when TS spectra are dominated by noise, or when fast analysis is required for real-time operation.  Although deep neural networks typically require large training sets, transfer learning can improve model performance on a target task with limited data by leveraging pre-trained models from related source tasks, where select hidden layers are further trained using target data.  We present five architecturally diverse deep neural networks, pre-trained on synthetic TS data and adapted for experimentally measured TS data, to evaluate the efficacy of transfer learning in estimating $n_e$ and $T_e$ in both the collective and non-collective scattering regimes.  We evaluate errors in $n_e$ and $T_e$ estimates as a function of training set size for models trained with and without transfer learning, and we observe decreases in model error from transfer learning when the training set contains $\lessapprox$ 200 experimentally measured spectra.
\end{abstract}
\maketitle
\section{Introduction}
Deep neural networks (DNNs), which can approximate any continuous function \cite{hornik_multilayer_1989}, have played a significant role in the advancement of various scientific fields \cite{nagpal_survey_2019}.  DNNs process data by passing numerical values through layers of interconnected nodes (or neurons), with each node transforming its input using a function defined by a set of trainable parameters, called weights and biases \cite{widrow_30_1990}.  DNNs offer a wide range of advantages, including fast processing \cite{mariscal_enhanced_2022} and robustness to noise \cite{rolnick_deep_2018}.  One limitation of DNNs is that they typically require large amounts of labeled training data.  This challenge motivates the need for models that can adapt to variable experimental parameters and learn sufficiently from small datasets \cite{dordevic_application_2023}.  An effective solution to this problem is transfer learning, a machine learning technique in which a model trained on a source task is adapted to a different, but related, target task \cite{bozinovski1976influence}.

An intuitive example of transfer learning is adapting a model trained to operate autonomous passenger cars for use in autonomous trucks.  With basic driving functions largely captured by the passenger car model, further training with a small set of truck driving data accounts for variations such as size and turning radius \cite{fent2024mantruckscenes}.  Transfer learning has already been applied to a wide range of plasma physics problems.  For example, a model trained on data from the J-TEXT tokamak was adapted to predict disruption events in the EAST tokamak, using data from only 20 EAST discharges \cite{zheng_disruption_2023}.  Inertial confinement fusion designs have been optimized to increase neutron yield using models that were pre-trained with simulation data \cite{humbird_transfer_2022}.  Plasma diagnostics have also benefited from transfer learning \cite{zhang_using_2023, falato_plasma_2022}, particularly in the context of data analysis for high-repetition-rate (HRR) experiments \cite{swanson_applications_2022, mariscal_design_2021, simpson_development_2021}.

In this paper, we explore several deep learning models for estimating $n_e$ and $T_e$ in a laser-produced plasma using Thomson scattering (TS) spectra obtained through HRR ($>1$ Hz) experiments at the PHOENIX Laser Laboratory \cite{pilgram_high_2023}.  Each model is pre-trained on synthetic TS data, spanning the collective and non-collective scattering regimes.  The synthetic non-collective spectra are generated using a Gaussian model that relates $n_e$ to the total signal intensity with a calibration factor obtained experimentally via Raman scattering \cite{kaloyan_raster_2021}.  The synthetic collective spectra are generated using PlasmaPy, a community-developed open source Python package that implements routines for simulating and fitting TS spectra \cite{plasmapy2025}.  Through transfer learning, the models are adapted to experimentally measured TS spectra, which involves additional training for select hidden layers near the output.  The ground truth $n_e$ and $T_e$ values for the experimental TS spectra are determined by a Gaussian fit for non-collective spectra and PlasmaPy's TS fitting procedure for collective spectra.  Following previous work that developed the laboratory's TS diagnostic \cite{zhang_two-dimensional_2023} and tested a neural network for analyzing TS spectra \cite{eisenbach_machine_2024}, we examine a diverse set of DNN architectures to establish a robust framework for identifying the optimal use cases of transfer learning for analyzing TS spectra.  We quantify the performance gain from transfer learning by evaluating the error on the test set for each type of DNN, with and without transfer learning, as a function of the training set size.  We also explore methods for quantifying model uncertainty through ensemble averaging and Bayesian neural networks.  This study is motivated by the challenges of HRR experiments, which benefit from rapid and accurate data analysis to inform driving algorithms between shots.

\section{Background}

\subsection{Deep Transfer Learning}\label{sec:back:dtl}
The foundational architecture of deep neural networks is the multi-layer perceptron (MLP), a feed-forward neural network with discrete layers, where each node in a given layer is connected to every node in the adjacent layers \cite{werbos1974beyond}.  The output of a node in an MLP is defined by the following equation:
\begin{equation}
a_i^{k} = \sigma\biggl(b_i^k+\sum_{j} w_{ij}^ka_j^{k-1}\biggr)
\end{equation}
where $a_i^{k}$ is the output of the $i$th node in layer $k$, $a_j^{k-1}$ is the output of the $j$th node in layer $k-1$, $w_{ij}^k$ is the weight of the connection between the two nodes, and $b_i^k$ is the bias of the $i$th node in layer $k$ \cite{haykin1999neural}.  The sum is the input for an activation function $\sigma$, which introduces nonlinearities and produces the final output of the node \cite{pmlr-v15-glorot11a}.

The core challenge of training neural networks is to find some combination of weights and biases that minimizes the average model error over a given dataset.  This average error is quantified by a cost function $C(\bm{\theta})$, where $\bm{\theta}$ is a vector representing the weights and biases of the model \cite{widrow1960adaptive}.  A standard algorithm for determining the values of the weights and biases that minimize the cost function is gradient descent \cite{bryson1969applied}.  This algorithm updates each weight and bias by repeating
\begin{equation}
\theta_{i}^{\,j+1}=\theta_{i}^{\,j}-\eta\frac{\partial C(\bm{\theta})}{\partial \theta_i}\Bigg|_{\bm{\theta}=\bm{\theta}^{j}}
\end{equation}
for every trainable parameter $\theta_i$ (where the superscript $j$ denotes the parameter values at the $j$th iteration) until a set number of iterations is reached or a convergence criterion is satisfied.  Here, $\eta$ is the learning rate, which determines the magnitude of the changes made to the weights and biases during each iteration \cite{robbins_stochastic_1951}.  All weights and biases are randomly initialized for the first iteration \cite{glorot_understanding_2010}.  For subsequent iterations, a forward pass is performed to compute the outputs $a_i^k$ for all nodes.  The output of the final layer is used to evaluate $C(\bm{\theta}^j)$, which quantifies the model error at the $j$th iteration.  This error is used to compute $\partial C(\bm{\theta})/\partial\theta_i$ for every $\theta_i$ via backpropagation, which recursively applies the chain rule from the output layer back to each trainable parameter \cite{rumelhart_learning_1986}.

A natural extension of this training process is deep transfer learning, a framework in which a deep neural network trained on a source task is adapted to a related target task \cite{bozinovski_reminder_2020}.  After initial training on the source dataset, the model parameters $\bm{\theta}$ are partitioned into frozen and trainable subsets, $\bm{\theta}_F$ and $\bm{\theta}_T$, respectively \cite{5288526}.  This partitioning preserves the full model architecture, with all layers and connections unchanged, and simply designates which parameters are trainable.  $\bm{\theta}_T$ is optimized via gradient descent using the target dataset, while $\bm{\theta}_F$ remains constant, though both subsets are used during forward passes to compute the target cost function.  $\bm{\theta}_T$ and $\bm{\theta}_F$ are often partitioned by layer, with entire layers designated as trainable or frozen.  Layers near the input are typically frozen, while layers closer to the output remain trainable.  This approach to transfer learning is effective because layers closer to the input capture general representations that are useful across many potential tasks.  In contrast, layers near the output adjust for features that are specific to the target task \cite{yosinski_how_2014}.  By preserving general representations in the early layers and allowing the later layers to specialize, the model adapts to the target task with minimal additional training data.

\subsection{Thomson scattering}

Thomson scattering (TS) describes the scattering of electromagnetic radiation by free charged particles \cite{froula2012plasma}.  Fundamental plasma parameters, such as $n_e$ and $T_e$, can be derived by analyzing the scattered power spectrum at wavelengths near the incident light.  When the incident electric field is polarized perpendicular to the scattering direction, the differential scattered power per unit frequency $\partial \omega$ per unit solid angle $\partial \Omega$ per unit volume $\partial V$ is defined by
\begin{equation}
\frac{\partial^3 P_s}{\partial \omega \, \partial \Omega \, \partial V}
= \frac{I_0 n_e r_0^2}{2\pi} \left( 1 + \frac{2\omega}{\omega_i} \right) S(\mathbf{k}, \omega)
\end{equation}
\noindent where $P_s$ is the total scattered power, $I_0$ is the incident laser intensity, $n_e$ is the electron density, $r_0$ is the classical electron radius, $\omega_i$ is the incident laser frequency, and $\omega=\omega_s-\omega_i$ where $\omega_s$ is the frequency of the scattered light \cite{lle2016plasma}.  $S(\mathbf{k},\omega)$ is the spectral density function, which determines the profile of the scattered power spectrum.  For plasmas consisting of electrons and multiple ion species, the spectral density function is defined as
\begin{equation}
\begin{aligned}
S(\mathbf{k},\omega) &= \frac{2\pi}{\left|\mathbf{k}\right|} \left|1 - \frac{\chi_e}{\epsilon} \right|^2 f_e\left(\frac{\omega}{\mathbf{k}}\right) \\[0.3ex]
&\quad + \sum_j \frac{2\pi}{\left|\mathbf{k}\right|} \frac{Z_j^2 n_j}{N} \left| \frac{\chi_j}{\epsilon} \right|^2f_j\left(\frac{\omega}{\mathbf{k}}\right)
\end{aligned}
\label{eqn:pectral density}
\end{equation}
where $\mathbf{k} = \mathbf{k}_s-\mathbf{k}_i$ is the scattering wave vector, $\mathbf{k}_i$ and $\mathbf{k}_s$ are the incident and scattered wave vectors, $\epsilon=1+\chi_e+\sum_j\chi_j$ is the plasma dielectric function, $\chi_e$ and $\chi_j$ are the electron and ion susceptibilities, $Z_j$ is the charge state of the $j$th ion population, $n_j$ is the density of the $j$th ion population, and $N$ is the combined density of all ion species \cite{foo_recovering_2023}.  $f_e(\omega/\mathbf{k})$ and $f_j(\omega/\mathbf{k})$ are the one-dimensional projections of the electron and ion velocity distributions in the direction of $\mathbf{k}$, which are assumed to be Maxwellian for both experimental and synthetic spectra \cite{pilgram_two-dimensional_2024}.

Thomson scattering occurs in the collective or non-collective scattering regime, depending on the scattering parameter $\alpha=1/(\left|\mathbf{k}\right|\lambda_{D})$, where $\lambda_{D}$ is the Debye wavelength.  In the collective regime ($\alpha>1$), the scattered power spectrum is characterized by scattering off plasma waves, where $n_e$ and $T_e$ are related to the wavelength separation and spectral width of the electron plasma wave features, respectively.  In the non-collective regime ($\alpha<1$), scattering off individual free charged particles dominates the spectrum, where $n_e$ and $T_e$ are related to the total signal intensity and spectral width, respectively.  This analysis incorporates data from both the collective and non-collective scattering regimes.

\subsection{Experimental Data}\label{sec:back:exp}

The experimental datasets used to train and test the models were produced during a study by Pilgram \textit{et al}. \cite{pilgram_high_2023} to investigate the generation of magnetic fields in Sedov-Taylor blast waves via the Biermann battery mechanism.  The same experimental design is described by Zhang \textit{et al}. \cite{zhang_two-dimensional_2023} and Eisenbach \textit{et al} \cite{eisenbach_machine_2024}.  The experiment involves irradiating a 38 mm diameter cylindrical high-density polyethylene (\ce{C2H4}) target at a repetition rate of 1 Hz with a high average power laser (10 J in 20 ns at 1053 nm) \cite{schaeffer_platform_2018}.  An $f/26$ spherical lens focuses the beam to a 250 $\mu$m focal spot on the target (yielding intensity on the order of $10^{12}$ W/cm$^2$) at an incident angle of 34$\degree$, which launches a laser-driven blast wave perpendicular to the target into the surrounding nitrogen gas (94 mTorr, 298 K).

The TS diagnostic for this experiment provides point measurements for $n_e$ and $T_e$ by scattering light from a laser pulse (460 mJ in 4 ns at 532 nm) off the laser-produced plasma.  The scattered light is captured by collection optics and focused into a fiber array, which is connected to a triple-grating spectrometer (TGS) \cite{ghazaryan_thomson_2022}.  A spatial notch filter (width of 1.7 $\pm$ 0.1 nm, centered at 532 nm) is placed in the light path inside the TGS for stray light suppression, and an image intensified camera (ICCD) at the output of the TGS acquires each TS spectrum.  The final recorded spectra contain 512 bins with 0.0388 nm/bin \cite{zhang_two-dimensional_2023}.

TS measurements are made over a planar region with coordinates $x \in \{-10, 0\}$ mm and $y \in \{4, 23\}$ mm relative to the target.  The scattering volume is translated with motorized mirrors in the TS beam path.  The collection optics, also mounted on stepper motors, move accordingly to maintain the alignment and scattering geometry, such that the incident unit wavevector is $\hat{k}_i = \hat{y}$ and the scattered unit wavevector is $\hat{k}_s = \hat{x}$.  Spectra at each coordinate are averaged over five individual shots and subtracted by five plasma-only background shots.  The experimental dataset consists of four separate planes of TS measurements, with each plane covering the same region in space and each spectrum measured at the same time $t=(100\pm5)$ ns after the drive laser.

Ground truth $n_e$ and $T_e$ values are obtained using two different methods depending on whether the scattering is collective or non-collective.  In the collective regime, PlasmaPy's TS fitting procedure is used to determine $n_e$ and $T_e$ by fitting a model of $S(\mathbf{k},\omega)$ to the measured spectrum \cite{zhang_two-dimensional_2023} using the lmfit Python package \cite{newville_2025_16175987}.  In the non-collective regime, $S(\mathbf{k},\omega)$ is effectively independent of $n_e$, so we model these spectral power distributions with a Gaussian function (see Section \ref{sec:back:syn} for details).  The total area under the fit is proportional to $n_e$, calibrated via Raman scattering, and $T_e$ is related to the Gaussian width.  $T_e$ can also be obtained using PlasmaPy's TS fitting procedure in the non-collective regime.

\subsection{Synthetic Data}\label{sec:back:syn}

The synthetic dataset contains 10,000 Thomson scattering spectral power distributions with corresponding $n_e$ and $T_e$ values, providing adequate resolution in $n_e$-$T_e$ parameter space with reasonable generation time ($<10$ s).  To produce the synthetic data, we use a Gaussian model for non-collective spectra and the PlasmaPy Thomson scattering spectral density function for collective spectra.  We use the Gaussian model to generate non-collective spectra because $n_e$ is proportional to the total signal intensity in this regime, which is not represented by PlasmaPy's model of $S(\mathbf{k},\omega)$.  In the collective regime, $n_e$ is related to the wavelength separation between the peaks of the electron plasma wave (EPW) features, which is encoded in the profile of the spectral density function.  $T_e$ depends only on the profile of the spectral density function in both scattering regimes, as it is related to the spectral width of the Gaussian distribution (non-collective) or EPW features (collective).

The Gaussian model for this experimental setup is derived using the relation between the measured total signal intensity $N_T$ and $n_e$.  The total signal intensity measured by the detector is defined by
\begin{equation}
N_T = \frac{I_i \tau_i}{h \nu_i} n_e \Delta V \frac{d\sigma_T}{d\Omega} \Delta \Omega\mu \zeta  G
\end{equation}
where $I_i$ is the incident laser intensity, $\tau_i$ is the incident laser pulse length, $h\nu_i$ is the energy of a single incident laser photon, $n_e$ is again the electron density, $\Delta V$ is the scattering volume, $d \sigma_T / d \Omega$ is the differential Thomson cross section, $\Delta \Omega$ is the scattering solid angle, $\mu$ is the transmission through the collection optics, $\zeta$ is the quantum efficiency of the ICCD, and $G$ is the gain of the ICCD \cite{ghazaryan_silica_2021}.  Eq. (3) simplifies to $n_e=s \cdot N_T$ because all terms except for $n_e$ are effectively constant throughout the experiment and common between the Thomson and Raman setups.  The calibration factor $s$ is quite difficult to calculate \textit{a priori}, so determining $s$ experimentally is often favored.  Using Raman scattering with nitrogen gas of known density (at a pressure of 0.88 bar), the calibration factor is determined to be $s=(3.11 \pm 0.22) \times 10^{10}$ cm$^{-3}$ / counts \cite{zhang_two-dimensional_2023}.

\begin{figure}[!ht]
    \centering
    \includegraphics[width=1\linewidth]{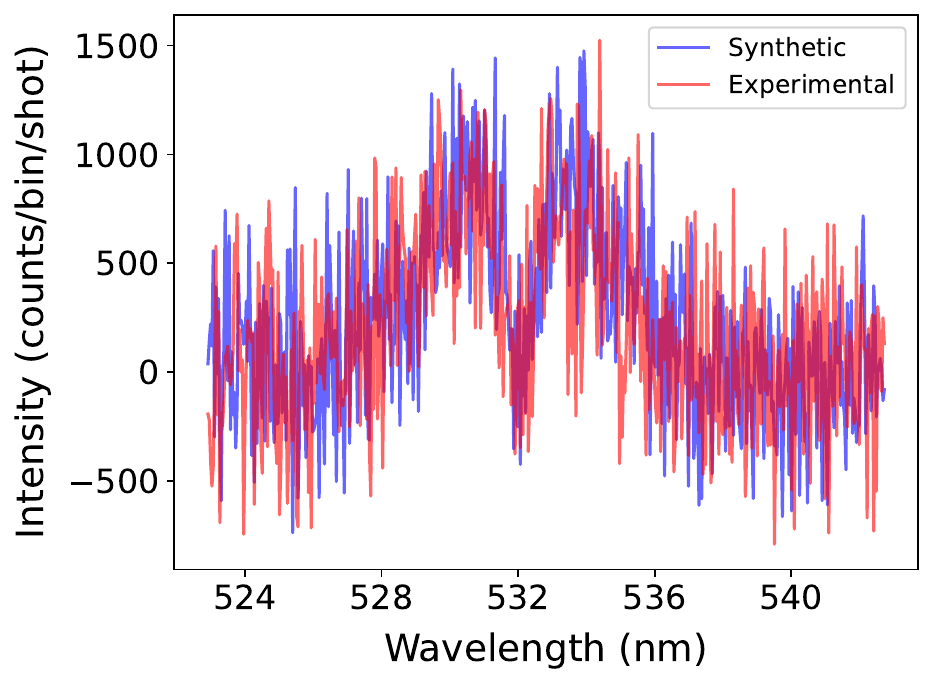}
    \caption{Comparison of synthetic (blue) and experimental (red) Thomson scattering spectra.  The synthetic spectrum is generated with the Gaussian model using the same values for electron density ($n_e=4.1 \times 10^{15}$ cm$^{-3}$) and electron temperature ($T_e=3.83$ eV) as those derived from a Gaussian fit to the experimental spectrum.  Negative values in the experimentally measured spectrum arise from background subtraction of plasma-only shots.}
    \label{fig:ExpSynCompare}
\end{figure}

Assuming the underlying electron velocity distribution is Maxwellian, we can model $N_T$ by
\begin{equation}
N_T=\int_{0}^{\infty} ae^{\textstyle -\frac{(\lambda_s - \lambda_i)^2}{2\sigma^2}} \, d\lambda
\label{eqn:gaussian-integral}
\end{equation}
where $a$ is the amplitude of the synthetic non-collective spectrum, $\lambda_s$ is the scattering wavelength, $\lambda_i=532$ nm is the probe wavelength, which is the center of the distribution, and $\sigma$ is the standard deviation.  This integral simplifies to $N_T=a \sigma \sqrt{2 \pi}$.  Using the relation derived from Raman scattering calibration, we can solve for the amplitude $a$ and obtain the following equation:
\begin{equation}
a = \frac{n_e}{s \sigma \sqrt{2 \pi}}
\label{eqn:gaussian-amplitude}
\end{equation}
We can also relate the standard deviation $\sigma$ to $T_e$ by
\begin{equation}
T_e = \frac{m_e c^2}{8 k_B} \left[ \frac{\Delta \lambda_{1/e}}{\lambda_i \sin\left(\frac{\theta}{2}\right)} \right]^2
\label{eqn:Te-spectral-halfwidth}
\end{equation}
where $m_e$ is the electron rest mass, $c$ is the speed of light, $k_B$ is the Boltzmann constant, and $\Delta \lambda_{1/e}$ is the spectral ($e^{-1}$) half-width, which is related to the standard deviation by $\Delta \lambda_{1/e} = \sigma \sqrt{2}$ \cite{kaloyan_first_2022}.  For probe wavelength $\lambda_i = 532$ nm and scattering angle $\theta = \frac{\pi}{2}$, Eq. (6) reduces to $T_e=0.903 \sigma^2$.  Solving for $\sigma$, we obtain our final expression for the Gaussian model of non-collective spectra for this experimental setup:
% f(\lambda_s)=\frac{n_ee^{- \textstyle \frac{0.903(\lambda_s - \lambda_i)^2}{2 T_e}}}{s \sqrt{\frac{2 \pi T_e}{0.903}}}
\begin{equation}
f(\lambda_s) = \frac{n_e}{s}\sqrt{\dfrac{0.903\hspace{0.1ex}}{2\pi  T_e}}
% \,e^{\frac{-0.903(\lambda_s-\lambda_i)^2}{2\hspace{0.1ex}T_e}}
\exp\left[-\frac{0.903\hspace{0.1ex} (\lambda_s - \lambda_i)^2}{2\hspace{0.1ex}T_e}\right]
\label{eqn:final-gaussian-model}
\end{equation}
where $f(\lambda_s)$ returns the expected measured intensity at $\lambda_s$ in units of counts/bin/shot given some $n_e$ and $T_e$.  To emulate the notch region in the experimental spectra, we multiply each synthetic non-collective spectrum by a Gaussian notch function \cite{eisenbach_machine_2024}, which is defined by
\begin{equation}
G = 1 - e^{\textstyle \frac{-(\lambda_s - \lambda_i)^2}{0.4608}}
\label{eqn:gaussian-notch}
\end{equation}
Random values drawn from a normal distribution ($\mu=0$, $\sigma=10^{2.5}$) are added to each synthetic non-collective spectrum for two purposes.  Primarily, injecting Gaussian noise to the inputs is equivalent to L2 regularization in the limit of infinitely many noise realizations \cite{bishop_training_1995}.  Although we only apply one noise array per spectrum, this approach still exposes the neural networks to perturbed inputs, partially reproducing the benefits of regularization via input noise.  Secondly, the synthetic noise amplitude (set by $\sigma$) is chosen to approximately match the signal-to-noise ratio (SNR) observed in the experimental spectra, particularly for low $n_e$ shots, without attempting to explicitly model all sources of noise.  Figure \ref{fig:ExpSynCompare} compares a synthetic and experimentally measured non-collective spectrum with the same corresponding $n_e$ and $T_e$.

Synthetic collective spectra are generated using the PlasmaPy spectral density function, which returns $S(\mathbf{k},\omega)$ from the scattering wavelength range, scattering geometry, probe wavelength, $n_e$, $T_e$, and ion temperature $T_i$.  All parameters are set to match the experimental setup described in Section \ref{sec:back:exp}.  We set $T_i=0.1$ eV across all synthetic collective spectra because its effect on $S(\mathbf{k}, \omega)$ is not relevant to the experimental data, where the ion acoustic wave (IAW) feature is suppressed by the notch filter.  To remove the IAW feature from the synthetic collective spectra, we set the values in the wavelength range between 531 nm and 533 nm to 0, since the Gaussian notch function distorts the IAW feature instead of removing it completely.  We add noise to the synthetic collective spectra using the same method as for the synthetic non-collective spectra, but with different parameters for the sampled normal distribution ($\mu = 0$ and $\sigma = 10^{-15}$).

\begin{figure}[!ht]
    \centering
    \includegraphics[width=1\linewidth]{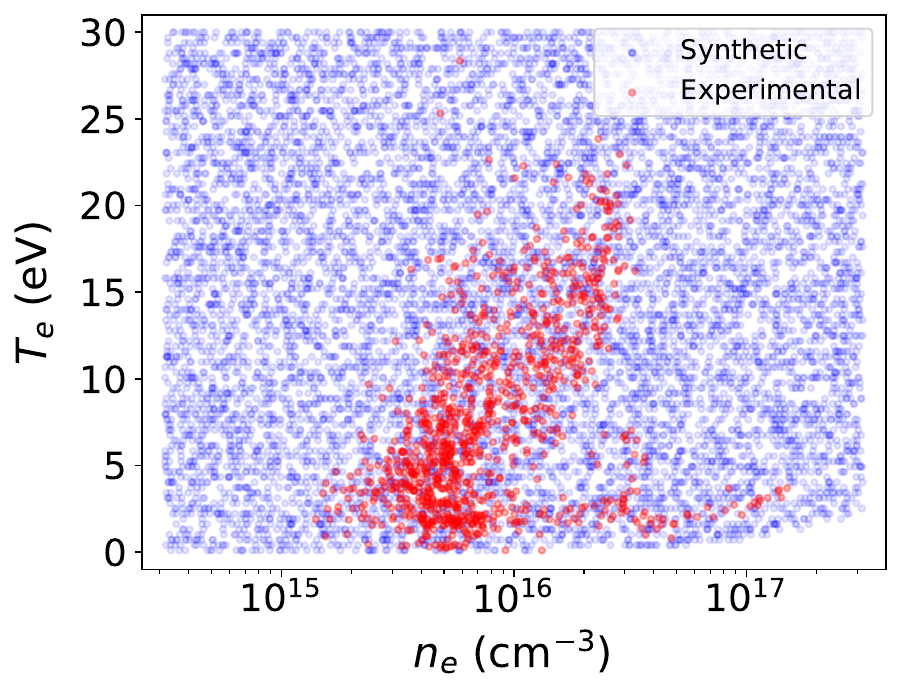}
    \caption{Distributions of synthetic (blue) and experimental (red) data across electron density and temperature.  While the experimental dataset is concentrated in a narrow region, the synthetic dataset covers a broader range of $n_e$ and $T_e$ values, which helps improve model generalization.  The constraint of $\alpha <3$ limits synthetic data from covering the lower-right region of the plot.}
    \label{fig:DataDistributions}
\end{figure}

To generate the complete synthetic dataset, we randomly sample $n_e$ and $T_e$ from the ranges [$10^{14.5}$, $10^{17.5}$] cm$^{-3}$ and [$0.1$, $30$] eV, respectively.  The $n_e$ values are sampled in logarithmic space, whereas the $T_e$ values are sampled in linear space.  If the randomly sampled combination of $n_e$ and $T_e$ results in $\alpha < 1$, these values are used to generate a synthetic spectrum with the Gaussian model.  If $1 \leq \alpha < 3$, these values are used to generate a spectrum with the PlasmaPy spectral density function.  To ensure the synthetic spectra are physically relevant to this experiment, we exclude combinations of $n_e$ and $T_e$ that lead to $\alpha > 3$.  This process is repeated until 10,000 synthetic spectra are generated.  Figure \ref{fig:DataDistributions} displays the distributions of the synthetic and experimental datasets in $n_e$–$T_e$ parameter space.  The synthetic dataset covers regions of the parameter space beyond what is experimentally measured for transferability to unseen experimental data, which is discussed by Eisenbach \textit{et al} \cite{eisenbach_machine_2024}.

\section{Methods}
\subsection{Data Pre-Processing}
The synthetic dataset is split with 67.5\% for the training set, 7.5\% for the validation set, and 25\% for the test set.  For the experimental dataset, three of the four planes (780 spectra) are used for the training/validation sets (90\%/10\% split), and the remaining plane (252 spectra) is used for the test set.  The inputs (spectral power distributions) and target values ($n_e$ and $T_e$) must be normalized before training to prevent numerical instabilities \cite{lecun2012efficient}.  Each synthetic and experimental spectrum is scaled by the maximum pixel value over the experimental training set so that the data fall in the range [0, 1].  The experimental spectra are also smoothed using a moving average with a window size of 10 pixels.  We apply log$_{10}$ scaling to the $n_e$ values to help prevent errors in $n_e$ estimates from having a disproportionate impact on the combined loss with $T_e$ during training.  This logarithmic transformation is applied across all models except for one, which will be discussed in Section \ref{sec:methods:setup}.  Then, the $n_e$ and $T_e$ values are scaled separately with min-max normalization:
\begin{equation}
x_{scaled}=\frac{x-x_{min}}{x_{max}-x_{min}}
\end{equation}
The model outputs for $n_e$ and $T_e$ are produced in normalized units, so we apply the inverse transformation to recover the original physical units.  Because the normalized model outputs are converted back to physical units, the test set values for $n_e$ and $T_e$ are kept in their original physical units for comparison and are not min-max normalized.

\subsection{Model Setups}\label{sec:methods:setup}
\begin{figure}[!ht]
    \centering
    \includegraphics[width=1\linewidth]{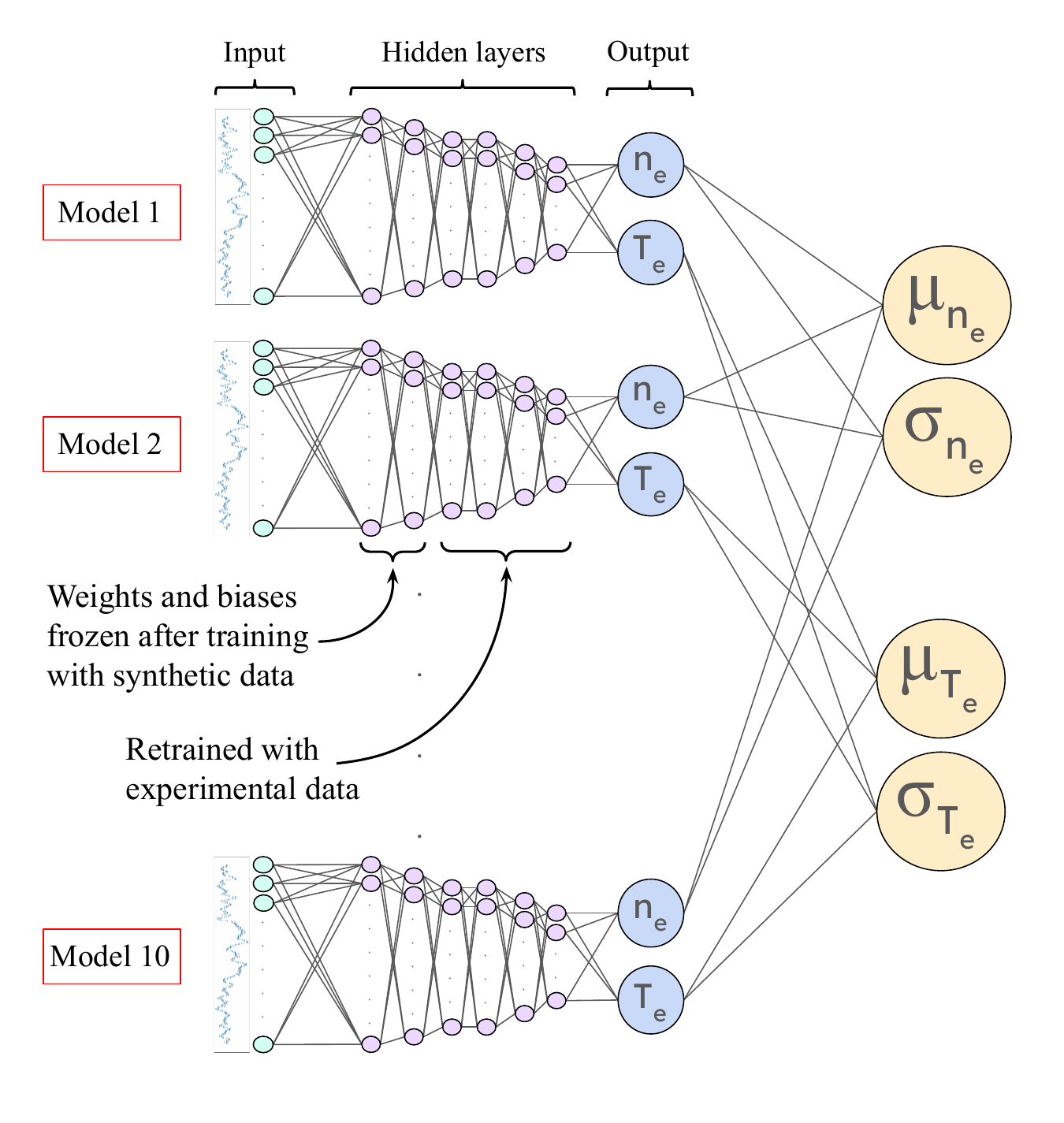}
    \caption{Diagram of ensemble averaging and transfer learning with the multi-layer perceptron.  The outputs of each model are combined to produce average estimates for $n_e$ and $T_e$ ($\mu_{n_e}$ and $\mu_{T_e}$) along with corresponding uncertainty estimates ($\sigma_{n_e}$ and $\sigma_{T_e}$).  }
    \label{fig:EnsembleMLP}
\end{figure}

All models were implemented with PyTorch 2.5.1 \cite{paszke_pytorch_2019} and trained on an NVIDIA GeForce RTX 4070 GPU.  The training process has two distinct steps: training with synthetic data and further training of select hidden layers with experimental data.  As discussed in Section \ref{sec:back:dtl}, all model architectures are preserved throughout the training process.  Here, we describe the five types of DNNs used to evaluate the efficacy of transfer learning for analyzing TS spectra.  These models include a standard multi-layer perceptron (MLP), a multi-branch MLP, a convolutional neural network (CNN), an ensemble MLP, and a Bayesian neural network (BNN).  For synthetic data pre-training, the models are optimized using mini-batch gradient descent with a batch size of 128 samples.  The rectified linear unit (ReLU) function, which is defined by $f(x)=$max$(0,x)$, is used for all activation functions.  Adaptive moment estimation (Adam) \cite{kingma_adam_2017} is used to accelerate and stabilize optimization by smoothing parameter updates to follow a more consistent path toward the cost function minimum.  We quantify the average difference between each model's estimates and the true values using the mean squared error (MSE) loss function, defined by
\begin{equation}
\mathcal{L}_{MSE}=\frac{1}{N}\sum_{i=1}^{N}(y_i-\hat{y_i})^2,
\end{equation}
where $N$ is the total number of values being estimated in a single mini-batch, $y_i$ is the model output for a given target value, and $\hat{y_i}$ is the corresponding ground truth value.  The range of models used here provides a thorough framework for studying transfer learning and its impact on model performance.
\newline
\noindent\textbf{1.\ Multi-Layer Perceptron}\vspace{3pt}\\
The simplest model that we train is an MLP with six hidden layers in a pyramidal structure, with layer sizes of 512, 256, 200, 200, 128, and 64 nodes.  The input layer consists of 512 nodes, corresponding to each point in our TS spectra, and the output layer consists of two nodes, which return $n_e$ and $T_e$.  The pyramidal structure gradually reduces the latent representation of the input, in contrast to a model with uniform layer sizes, allowing the model to extract increasingly compact features \cite{goodfellow2016deep}.  Each layer is fully connected to adjacent layers, providing direct integration of all input features to extract global patterns in the spectra.
\vspace{8pt}
\newline
\noindent\textbf{2.\ Multi-Branch MLP}\vspace{3pt}\\
Building on the MLP, we construct a multi-branch MLP in which the input layer splits into two independent branches \cite{covington_deep_2016}.  Each branch mirrors the original MLP's structure, with the same number of hidden layers and the corresponding number of nodes in each layer.  A multi-branch MLP is appropriate for this study because $n_e$ and $T_e$ are predominantly determined by separate spectral features.  The multi-branch MLP is designed to exploit this, with each branch optimized for features most relevant to its target.  One branch of the model adjusts its weights and biases specifically for $n_e$ and the other for $T_e$.  A key advantage of estimating $n_e$ and $T_e$ separately is that we can avoid log$_{10}$ scaling $n_e$ values because $n_e$ loss and $T_e$ loss are calculated separately.  Spectral power varies linearly with $n_e$, so bypassing log$_{10}$ scaling preserves this linear relationship and avoids introducing an artificial logarithmic dependence.
\vspace{8pt}
\newline
\noindent\textbf{3.\ Convolutional Neural Network}\vspace{3pt}\\
We evaluate the performance of a convolutional neural network (CNN) \cite{lecun_gradient-based_1998}, which differs from an MLP in that the nodes of adjacent layers are sparsely connected rather than fully connected.  Each node in a particular layer of a CNN is connected only to a local region of nodes in the adjacent layers, which allows CNNs to infer spatial properties of the input data \cite{simonyan_very_2015}.  CNNs are well-suited for analyzing TS spectra because $n_e$ and $T_e$ are correlated with local features of the spectral power distribution.  The CNN used in this study consists of four 1D convolutional layers with increasing channel depths of 64, 128, 256, and 512, each using a kernel size of 3 wavelength bins and a padding of 1 bin.  These layers are followed by a 1D max pooling layer with a kernel size and stride of 2 wavelength bins.  The output of the pooling layer is passed through a sequence of four hidden dense layers with 512, 256, 128, and 64 nodes, respectively.  The output layer again consists of two nodes for $n_e$ and $T_e$.
\vspace{8pt}
\newline
\noindent\textbf{4.\ Ensemble MLP}\vspace{3pt}\\
Assessing the confidence in a DNN's outputs is useful for interpretability \cite{goodfellow2016deep}.  Model estimates with high error can still be informative if they are accompanied by quantifiable uncertainty.  For instance, high model uncertainty may indicate that the input spectrum has a low SNR or has features that are not represented in the training set \cite{lakshminarayanan_simple_2017}.  For all of the models presented so far, we can use ensemble averaging to quantify model uncertainty.  Ensemble averaging combines several models such that the final output is the mean of the individual model outputs \cite{dietterich_ensemble_2000}.  Each model in the ensemble can have a distinct set of hyperparameters, and they can all be trained on different datasets.  Even when all models are trained on the same dataset with identical hyperparameters, each model is assigned a different random seed, producing a unique set of initial trainable parameters. These variations in initialization can lead to slightly different convergence paths during training, causing minor deviations in the final outputs of each model.  The upper and lower bounds of the ensemble output reflect the model's uncertainty.  Figure \ref{fig:EnsembleMLP} depicts the integration of ensemble averaging with transfer learning for a group of MLPs.  The ensemble size is a tunable parameter, and we use 10 models in this analysis.  Although not illustrated, the MLP, CNN, and ensemble MLP output the scattering parameter $\alpha$ as an auxiliary task, improving $n_e$ and $T_e$ estimates.

\noindent\textbf{5.\ Bayesian Neural Network}\vspace{3pt}\\
Model uncertainty can also be quantified by using probabilistic neural networks \cite{specht1990probabilistic}.  In contrast to ensemble averaging, probabilistic neural networks provide uncertainty measurements from a single model, offering greater computational efficiency.  Our choice of probabilistic neural network for this study is a Bayesian neural network (BNN) \cite{mackay_practical_1992}.  Unlike the models discussed previously, the weights and biases of a BNN are modeled as probability distributions rather than deterministic values.  The uncertainty in all of the weights and biases determines the uncertainty in the model's final outputs.  To train a BNN, the probability distributions of the trainable parameters are initialized by prior distributions \cite{jones_improving_2024}, which we choose to be Gaussian distributions for all weights and biases.  Gradient descent and backpropagation are still used to update the distributions of the weights and biases.  However, an additional loss term, the Kullback-Leibler (KL) divergence \cite{kullback1951information}, is introduced to penalize the model for deviating far from the prior distributions.  KL divergence is defined by
\begin{equation}
\mathcal{L}_{KL}=\frac{1}{M}\sum_{j=1}^M\mathrm{ln}\bigg(\frac{\sigma_{prior}}{\sigma_j}\bigg)+\frac{\sigma_j^2+(\mu_j-\mu_{prior})^2}{2\sigma_{prior}^2}-\frac{1}{2}
\end{equation}
where $M$ is the total number of trainable parameters, which includes the mean and standard deviation of each weight and bias distribution.  $\mu_j$ and $\sigma_j$ are the mean and standard deviation for the $j$th weight or bias, while $\mu_{prior}$ and $\sigma_{prior}$ are the mean and standard deviation of the prior distribution associated with the $j$th weight or bias \cite{blundell_weight_2015}.  When $\mu_j=\mu_{prior}$ and $\sigma_j=\sigma_{prior}$, $\mathcal{L}_{KL}=0$.  We choose $\mu_{prior}=0$ and $\sigma_{prior}=0.1$ for every weight and bias.  Our total loss function for the BNN is $\mathcal{L}_{total}=\mathcal{L}_{MSE}+\beta\mathcal{L}_{KL}$ where $\beta$ is a tunable parameter that determines the contribution of $\mathcal{L}_{KL}$ to $\mathcal{L}_{total}$.  We trained multiple models with different values of $\beta$ and evaluated their performance on the validation set.  The model with $\beta=0.1$ yielded the lowest validation MSE for both $n_e$ and $T_e$ estimates, and we use $\beta=0.1$ for all results presented here.  The BNN architecture consists of four hidden Bayesian linear layers with 512, 256, 128, and 64 nodes, decreasing in size along the feed-forward direction.  The output is a Bayesian linear layer with two nodes for $n_e$ and $T_e$.

\subsection{Hyperparameter Tuning}\label{sec:methods:bo}

\begin{figure*}
    \centering
    \raisebox{0.5cm}{\includegraphics[width=1\linewidth]{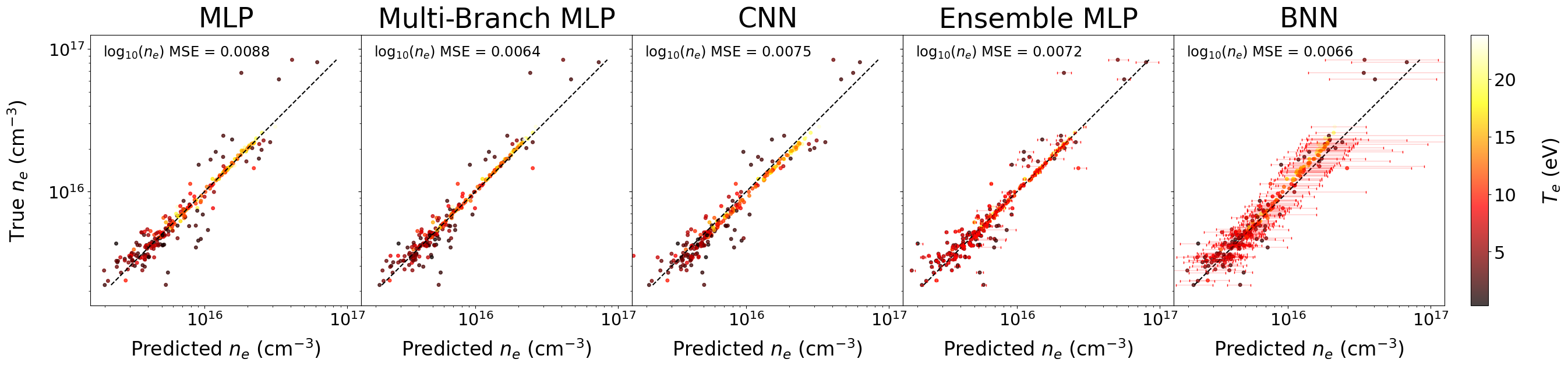}}
    % \vspace{0.001cm}
    \includegraphics[width=1\linewidth]{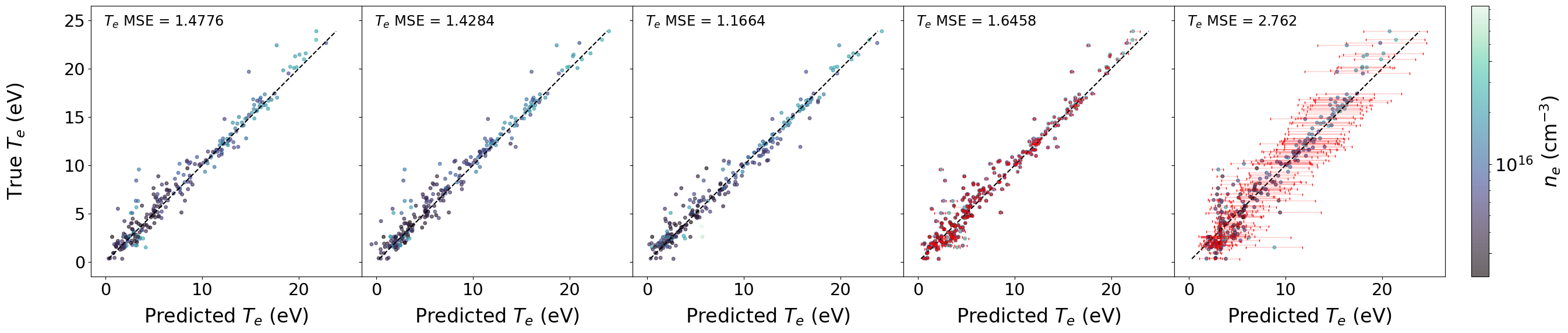}
    \caption{Model estimates versus ground truth values for electron density (top row) and electron temperature (bottom row).  Columns are organized by neural network architecture.  Points in the top row are shaded according to the corresponding $T_e$, while points in the bottom row are shaded according to the corresponding $n_e$.  Error bars span the range of model estimates for the ensemble MLP and BNN.  The dashed lines represent where model estimates would fall under perfect agreement with the ground truth values.}
    \label{fig:accuracy_plots}
\end{figure*}

Determining the optimal set of hyperparameters can lead to improvements in model estimates of $n_e$ and $T_e$.  While there are many hyperparameters and performance metrics to consider, we focus on finding the combination of learning rate and number of training epochs that minimizes the MSE on the validation set during synthetic data pre-training.  We represent a combination of learning rate and number of training epochs by the parameter vector $\mathbf{x}$.  Calculating the validation MSE for some $\mathbf{x}$ is computationally expensive because it requires training the model with its associated learning rate and number of training epochs.  To efficiently search this large parameter space, we use Bayesian optimization \cite{snoek_practical_2012} to test values of $\mathbf{x}$ that are most likely to yield a validation MSE less than the known minimum.  Rather than computing the validation MSE for every candidate $\mathbf{x}$ in a grid search, we use a Gaussian process to estimate a probability distribution $p(y \mid \mathbf{x})$ over the possible MSE values $y$ for a given $\mathbf{x}$.  We model the likelihood that some $\mathbf{x}$ will yield a validation MSE less than the known minimum with the expected improvement acquisition function \cite{jones_efficient_1998}:
\begin{equation}
    \mathrm{EI}(\mathbf{x}) = \int_{0}^{f^*} \left( f^* - y \right) p(y \mid \mathbf{x}) \, dy
\end{equation}
where $f^*$ is the current lowest validation MSE.  The integration limits restrict the calculation to the region of $p(y \mid \mathbf{x})$ that corresponds to potential improvements in validation MSE.  The expected improvement function balances exploration and exploitation.  When the mean of $p(y \mid \mathbf{x})$ is small, a significant region of the distribution lies within the integration limits.  Alternatively, if the predicted mean is large, a significant region of the distribution can still lie within the integration limits if the variance of $p(y \mid \mathbf{x})$ is large.  For points near the best-known $\mathbf{x}$, the predicted mean is close to $f^*$, and points in sparsely explored regions are associated with increased variance.  The term $\left( f^* \! - y \right)$ increases $\mathrm{EI}(\mathbf{x})$ for regions where greater improvements in validation MSE are expected.  The value of $\mathbf{x}$ that maximizes $\mathrm{EI}(\mathbf{x})$ is used to train a model in order to compute the validation MSE.

We implement Bayesian optimization with the python-based Ax library \cite{olson_ax_2025}, which is built on the BoTorch \cite{balandat_botorch_2020} library.  The algorithm begins by manual selection of an initial combination of learning rate and number of epochs $\mathbf{x}$ to train a model and compute the validation MSE.  The next five values for $\mathbf{x}$ are selected via Sobol sampling, which provides a more uniform coverage of the parameter space than random sampling \cite{bratley_algorithm_1988}.  The subsequent iterations select the value of $\mathbf{x}$ that maximizes $\mathrm{EI}(\mathbf{x})$ to train a model and compute the validation MSE.

\renewcommand{\arraystretch}{1.5}
\setlength{\tabcolsep}{6.5pt}

\begin{table}[h!]
\centering
\begin{tabular}{c|c|c}
\textbf{Model} & \textbf{Learning Rate} & \textbf{Epochs} \\
\hline
\hline
MLP & $1.301 \times 10^{-3}$ & 3169 \\
\hline
Multi-Branch MLP & $2.495 \times 10^{-4}$ & 2548 \\
\hline
CNN & $1.287 \times 10^{-6}$ & 4542 \\
\hline
Ensemble MLP & $1.301 \times 10^{-3}$ & 3169 \\
\hline
BNN & $4.982 \times 10^{-3}$ & 3638 \\
\end{tabular}
\caption{Best combination of learning rate and number of training epochs for each model during synthetic data pre-training, determined by Bayesian optimization over 100 iterations per model.  Random seeds were varied at each iteration to identify hyperparameters that are optimal across different model initializations.  The ensemble MLP inherits the best-known hyperparameters for the MLP.  The initial learning rate was set to $7.5 \times 10^{-5}$ and the initial number of training epochs to 1000 for all models.  The search space covered learning rates from $10^{-6}$ to 0.1 and training epochs from 500 to 5000.}
\label{tab:thomson_params}
\end{table}

\subsection{Transfer Learning Procedure}
After pre-training each model on synthetic TS data, the model parameters of the hidden layers closest to the input are frozen.  For the multi-layer perceptron, the parameters of the first two layers adjacent to the input are frozen.  The same strategy is applied to the multi-branch MLP, where the first two hidden layers in both the $n_e$ and $T_e$ branches are frozen.  Similarly, the first two layers of each model in the ensemble MLP are frozen.  For the convolutional neural network, the first two 1D convolutional layers are frozen.  Lastly, for the Bayesian neural network, the first two Bayesian linear layers are frozen.

To maintain consistency when comparing DNN performance, hyperparameters are kept identical across all models during training on experimental data, as applicable.  Consistent with pre-training on synthetic data, ReLU is used for all activation functions.  Each model, except for the BNN, is initialized with a learning rate of $7.5 \times 10^{-5}$, and the trainable parameters are updated using mini-batch gradient descent with a batch size of 128 samples over 1000 epochs.  The BNN is initialized with a learning rate of 0.01 and trained using the entire experimental training set as a single batch over 5000 epochs.  All models are trained with the Adam optimizer and the MSE loss function used during pre-training on synthetic data, and the BNN loss function is still augmented by the same weighted KL divergence term added to the MSE loss.

\section{Results}
All five models are evaluated using the same test set of TS spectra collected over a planar region surrounding the Sedov-Taylor blast wave, as described in Section \ref{sec:back:exp}.  The two main goals of this assessment are to evaluate the accuracy of each DNN and identify the regimes in which transfer learning improves model accuracy.

\begin{figure}
    \centering
    \includegraphics[width=1\linewidth]{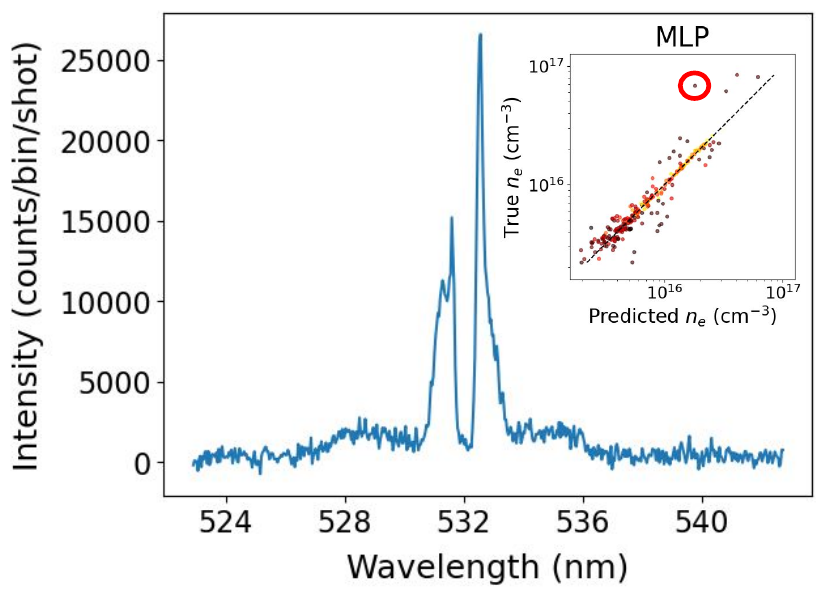}
    \caption{Example TS spectrum corrupted by stray light, revealed by the spike in intensity near the probe wavelength.  \textbf{Inset}: MLP $n_e$ estimates (x-axis) vs. ground truth values (y-axis), with the point circled in red corresponding to the example spectrum.  All models underestimate $n_e$ for this spectrum.}
    \label{fig:straylightfig}
\end{figure}

After pre-training on synthetic data, we train each model on the entire experimental training set and evaluate the accuracy of the $n_e$ and $T_e$ estimates relative to the ground truth values obtained using the Gaussian fit (non-collective spectra) and PlasmaPy’s TS fitting procedure (collective spectra).  Figure \ref{fig:accuracy_plots} displays model estimates versus ground truth values separately for $n_e$ and $T_e$.  Among the models evaluated, the multi-branch MLP achieves the lowest log$_{10}(n_e)$ MSE, while the standard MLP produces the highest log$_{10}(n_e)$ MSE.  The convolutional neural network achieves the lowest $T_e$ MSE, while the Bayesian neural network produces the highest $T_e$ MSE.  All models produce high errors in $n_e$ estimates when the spectra are corrupted by stray light, as shown by Fig.~\ref{fig:straylightfig}.  The Bayesian neural network produces the highest uncertainties in $n_e$ estimates for spectra corrupted by stray light.

\begin{figure}
    \centering
    \includegraphics[width=1\linewidth]{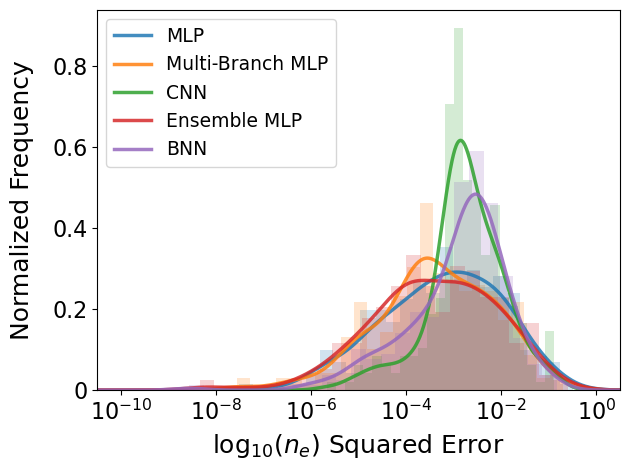}
    \includegraphics[width=1\linewidth]{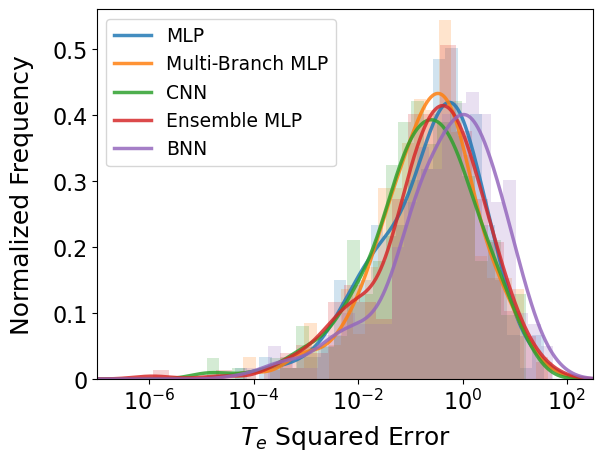}
    \caption{Distributions of $n_e$ and $T_e$ squared error for each model across all test spectra.  The histograms are normalized and smoothed with kernel density estimates (solid lines) \cite{silverman1986density}.  The logarithmic x-axes highlight substantial variability in model accuracy for both $n_e$ and $T_e$ estimates.}
    \label{fig:histogram}
\end{figure}

While MSE provides an overall measure of performance \cite{chicco_coefficient_2021}, analyzing the distribution of individual squared errors offers a more detailed picture of expected model accuracy \cite{gneiting_strictly_2007}.  Figure \ref{fig:histogram} displays the distributions of $n_e$ squared error and $T_e$ squared error for all five models.  All distributions in Fig.~\ref{fig:histogram} are plotted with the x-axis on a log$_{10}$ scale for visual convenience.  The convolutional neural network presents the narrowest distribution for $n_e$ squared error, which reflects its consistent performance across the entire test set.  By contrast, the $n_e$ squared error distributions for the three MLP variants are relatively wide.  All $T_e$ squared error distributions exhibit a slight negative skew, indicating that the models estimate $T_e$ more accurately for a small subset of samples compared to the broader test set.  The distributions span several orders of magnitude for all models, revealing the sensitivity of model accuracy to the input spectrum.

After analyzing the models trained with transfer learning, we next compare their performance to models trained without transfer learning.  Such comparisons help identify the conditions under which transfer learning improves model accuracy.  We therefore train five additional models using the same hyperparameters and architecture as the original five models.  Unlike the original models, which were first trained on synthetic data and fine-tuned with experimental data, these new models are trained directly on experimental data.  To quantify the regimes where transfer learning improves model performance, we compare the MSE on the test set between models trained with and without transfer learning as a function of the experimental training set size.  Each model is trained on progressively larger subsets of the experimental training data, and at each step, we compute log$_{10}(n_e)$ MSE and $T_e$ MSE on the test set.  The subset size begins at one sample and increases by one sample until reaching the full set of 780 samples.  We plot the results for each model separately, with individual plots for log$_{10}(n_e)$ MSE and $T_e$ MSE.

\begin{figure*}
    \centering
    \includegraphics[width=1\linewidth]{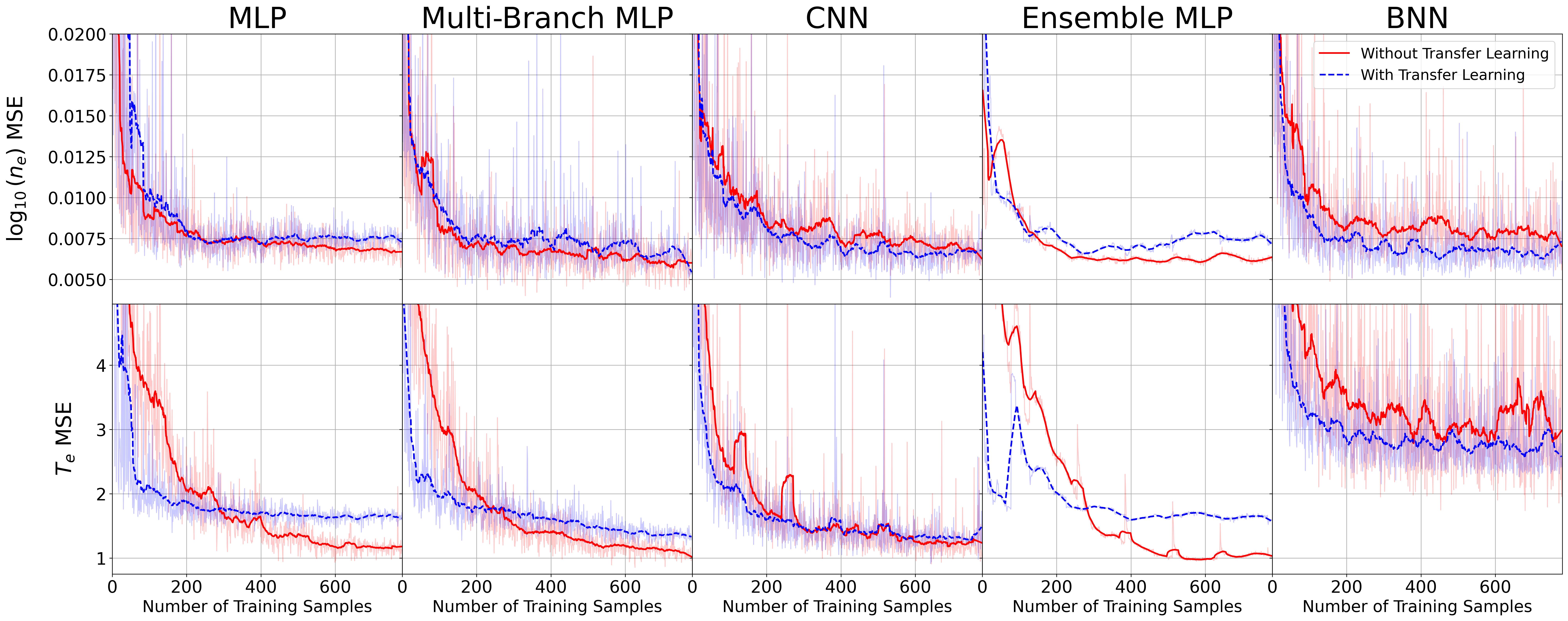}
    \caption{Model error on the test set as a function of training set size for each deep neural network trained with transfer learning (blue) and without transfer learning (red).  The number of training samples increases with a step size of one, and the random seed is constant across each iteration.  The curves are smoothed with a Savitzky-Golay filter \cite{savitzky_smoothing_1964} (window size = 31 points, first-order polynomial) and overlaid on the raw data.  The top row plots MSE for log$_{10}$-scaled $n_e$ estimates, and the bottom row plots MSE for $T_e$ estimates.  Columns are ordered by neural network architecture.  When the training set contains a small number of experimentally measured spectra, transfer learning generally improves model accuracy, especially for $T_e$ estimates.  While $n_e$ MSE levels off, $T_e$ MSE continues to decline for the multi-branch MLP, even with the full set of experimental training data.  More experimental training data may therefore improve $T_e$ estimates.}
    \label{fig:SamplesVSMSE}
\end{figure*}

Figure \ref{fig:SamplesVSMSE} demonstrates that models trained without transfer learning require approximately 200 experimentally measured spectra to converge to the performance of models trained with transfer learning.  Model improvements from transfer learning are most notable for $T_e$ estimates compared to $n_e$ estimates.  Periodic spikes in error are noticeable, especially for the ensemble MLP.  These spikes occur at multiples of 128, which is the batch size used to train each model (except the BNN). As the training subset size increases, new batches are formed at these multiples. Just after each multiple is reached, the newly created batch contains just a small number of training samples.  This causes the model to develop a bias toward those few samples, which negatively affects generalization and leads to poor estimates \cite{masters_revisiting_2018}.  For a small number of training samples ($\lessapprox$ 200), the multi-branch MLP shows a substantial reduction in $T_e$ MSE as a result of transfer learning and achieves low overall log$_{10}(n_e)$ MSE and $T_e$ MSE.  These results motivate a closer investigation of the multi-branch MLP, particularly when the experimental training set contains few samples.

\begin{figure*}
    \centering
    \includegraphics[width=1\linewidth]{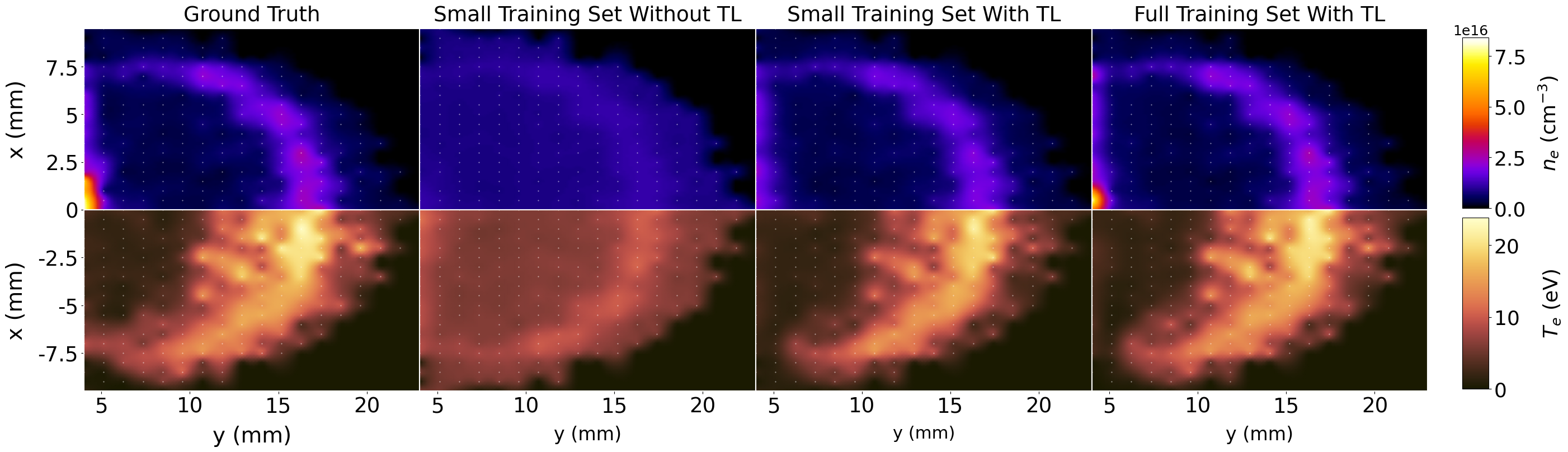}
    \caption{Images of Sedov-Taylor blast waves from $n_e$ (top row) and $T_e$ (bottom row) heat maps comparing ground truth values to model estimates.  The bottom row is mirrored to negative $x$ positions for convenient visual comparison.  White points indicate where the SNR was sufficient to compute $n_e$ and $T_e$ ground truth values.  For low SNR shots, $n_e$ and $T_e$ are interpolated within the blast wave region and set to zero in the upstream.  The first column displays the ground truth values of $n_e$ and $T_e$ obtained from Gaussian fits (non-collective spectra) and PlasmaPy's TS fitting procedure (collective spectra).  The second column displays $n_e$ and $T_e$ estimates from the multi-branch MLP trained without transfer learning (TL) on 10 experimental spectra, while the third column displays estimates from the multi-branch MLP trained with TL on the same 10 experimental spectra.  The last column displays the multi-branch MLP estimates trained with TL on the entire experimental training set.}
    \label{fig:BlastWave}
\end{figure*}

We next examine the multi-branch MLP’s ability to construct an image of the Sedov-Taylor blast wave using $n_e$ and $T_e$ heat maps, with a particular interest in the differences with and without transfer learning for a small experimental training set.  To aid in the comparisons, Fig.~\ref{fig:BlastWave} includes the blast wave images from the test set ground truth values and from the multi-branch MLP trained on the full experimental training set with transfer learning.  The small experimental training set contains only 10 spectra, and the models trained on this subset are limited to 100 training epochs to mitigate overfitting.  While the model trained without transfer learning is able to weakly construct the shape of the blast wave, it is not able to reasonably estimate the magnitudes of $n_e$ and $T_e$.  The images from the model trained with transfer learning on the small experimental training set match well with the ground truth values, except for a notable underestimation of $n_e$, especially near the target.  Training the model on the full experimental training set diminishes these errors.

\section{Conclusions}
Several deep neural network architectures are evaluated by their ability to accurately estimate $n_e$ and $T_e$ from collective and non-collective Thomson scattering spectra.  Through transfer learning, models trained on synthetic data are adapted to analyze experimental data and produce accurate results, even with minimal experimental training data.  While the synthetic collective data are generated with PlasmaPy’s spectral density function, the synthetic non-collective data must be generated with a Gaussian model that relates the total signal intensity to $n_e$ via Raman scattering calibration.  Models that are pre-trained with synthetic data outperform those trained only on experimental data when the experimental training set contains $\lessapprox$ 200 samples.  The margin of improvement decreases as the experimental training set grows, and eventually, models trained only on experimental data produce more accurate $n_e$ and $T_e$ estimates.  The ensemble MLP and Bayesian neural network quantify model uncertainty for both $n_e$ and $T_e$ estimates, providing an associated measure of reliability for each output.  All models produce $n_e$ and $T_e$ estimates on millisecond timescales, enabling analysis between shots for HRR experiments.  The MLP is implemented with the real-time Thomson scattering diagnostic at the PHOENIX Laser Laboratory, a LaserNetUS facility located on the UCLA campus.

\section{Acknowledgments}
This work was performed under the auspices of the U.S. Department of Energy by Lawrence Livermore National Laboratory under Contract DE-AC52-07NA27344 and funded by the LLNL LDRD program under tracking code 23-ERD-035. This work was supported by the Department of Energy (DOE) under award number DE-SC0024549, the Defense Threat Reduction Agency (DTRA) and Livermore National Laboratory under contract number B661613, the National Nuclear Security Administration (NNSA) Center for Matter Under Extreme Conditions under Award Number DE-NA0004147, the Naval Information Warfare Center-Pacific (NIWC) under contract NCRADA-NIWCPacific-19-354, the Department of Energy National Nuclear Security Administration under Award Numbers DE-NA0003856, DE-SC0020431, and DE-NA0004033, the University of Rochester, and the New York State Energy Research and Development Authority.

\section{Data Availability}
The data that support the findings of this study are available from the corresponding author upon reasonable request.

%\bibliographystyle{apsrev4-2}
%\bibliography{references}


\end{document}